\documentstyle[preprint,aps,epsfig]{revtex}

\begin{document}
\draft
\preprint{\begin{tabular}{l}
\hbox to\hsize{April, 1999 \hfill KAIST-TH 99/1}\\[-3mm]
\hbox to\hsize{hep-ph/9904283 \hfill SNUTP 99-003}\\[5mm] \end{tabular} }

\bigskip

\title{ 
Effects of supersymmetric  CP violating phases \\ 
on  $B \rightarrow X_s l^+ l^-$ and $\epsilon_K$ }
\author{Seungwon Baek\footnote{swbaek@muon.kaist.ac.kr} 
and Pyungwon Ko\footnote{pko@muon.kaist.ac.kr} }
\address{Department of Physics, KAIST \\ Taejon 305-701, Korea}
\maketitle
\begin{abstract}
We consider effects of  CP violating phases in $\mu$ and $A_t$ parameters in 
the effective supersymmetric standard model on $B\rightarrow X_s l^+ l^-$ 
and $\epsilon_K$. Scanning over the MSSM parameter space with experimental 
constraints including edm constraints from Chang-Keung-Pilaftsis (CKP) 
mechanism, we find that the ${\rm Br} (B\rightarrow X_s l^+ l^-)$ can be 
enhanced by upto $\sim 85 \%$ compared to the standard model (SM) prediction, 
and its correlation with ${\rm Br} ( B \rightarrow X_s \gamma)$ is distinctly 
different from the minimal supergravity scenario. Also we find $1 \lesssim 
\epsilon_K / \epsilon_K^{SM} \lesssim 1.4$, and fully supersymmetric CP 
violation in $K_L \rightarrow  \pi \pi$ is not possible. Namely, 
$|\epsilon_K^{\rm SUSY}| \lesssim O(10^{-5})$ if the phases of $\mu$
and $A_t$ are the sole origin of CP violation.
\end{abstract}


\newpage
\narrowtext

{\bf 1}. 
In the minimal supersymmetric standard model (MSSM), there can be many new
CP violating (CPV) phases beyond the KM phase in the standard model (SM)
both in the flavor conserving and flavor violating sectors.
The flavor conserving CPV phases in the MSSM are strongly constrained by 
electron/neutron electric dipole moment (edm) and believed to be very small 
($\delta \lesssim 10^{-2}$ for $M_{\rm SUSY} \sim O(100)$ GeV ) \cite{susycp}. 
Or, one can imagine that the 1st/2nd generation scalar fermions are very
heavy so that edm constraints are evaded via decoupling even for CPV phases
of order $O(1)$ \cite{kaplan}. 
Also it is possible that various contributions to 
electron/neutron EDM  cancel with each other in substantial parts of the 
MSSM parameter space even if SUSY CPV phases are  $\sim O(1)$ and SUSY 
particles are relatively light \cite{nath} \cite{kane}.
In the last two cases where SUSY CPV phases are of $\sim O(1)$, these 
phases may affect $B$ and $K$ physics in various manners. 
In the previous letter \cite{baek}, we presented effects of these SUSY CPV 
phases on $B$ physics : the $B^0 - \overline{B^0}$ mixing and the direct 
asymmetry in $B\rightarrow X_s \gamma$, assuming that EDM constraints and 
SUSY FCNC problems are evaded by heavy 1st/2nd generation scalar fermions. 
In this letter, we extend our previous work to $B\rightarrow X_s l^+ l^-$
and $\epsilon_K$ (see also Ref.~\cite{demir}.) within the same assumptions.

An important ingredient for large $\tan\beta$ in our model is the constraint 
on the $\mu$ and $A_t$ phases coming from electron/neutron edm's through 
Chang-Keung-Pilaftsis (CKP)  mechanism \cite{chang}.
Two loop diagrams with CP-odd higgs and photon (gluon) exchanges between
the fermion line and the sfermion loop (mainly stops and sbottoms) can 
contribute significantly to electron/neutron edm's in the large $\tan\beta$
region.  The authors of Ref.~\cite{chang}  find that 
\begin{equation}
( {d_f \over e } )_{\rm CKP} = Q_f {3 \alpha_{\rm em} \over 64 \pi^2}
{R_f ~m_f \over M_A^2} \sum_{q=t,b}~\xi_q Q_q^2 \left[ F
\left( { M_{\tilde{q}_1}^2 \over M_A^2 } \right) - F  
\left( { M_{\tilde{q}_2}^2 \over M_A^2 } \right) \right],
\end{equation} 
where $R_f = \cot\beta (\tan\beta)$ for $I_{3 f} = 1/2 ~(-1/2)$, and 
\begin{equation}
\xi_t = {\sin 2\theta_{\tilde{t}} m_t {\rm Im} (\mu e^{i \delta_t} ) 
\over \sin^2 \beta ~v^2}, ~~~~
\xi_b = {\sin 2\theta_{\tilde{b}} m_b {\rm Im} ( A_b e^{-i \delta_b} ) 
\over \sin \beta ~\cos\beta~v^2},
\end{equation}
with $\delta_q = {\rm Arg} (A_q + R_q \mu^* )$, and $F(z)$ is a two-loop
function given in Ref.~\cite{chang}.  This new contribution is independent
of the 1st/2nd generation scalar fermion masses, so that it does
not decouple for heavy 1st/2nd generation scalar fermions. Therefore it
can be important for the electron or down quark edm for the large $\tan\beta$ 
case. This is in sharp contrast with the usual one-loop contributions to 
edm's, for which \cite{chang} 
\begin{equation}
\left( {d_f \over e} \right) \sim 10^{-25} {\rm cm} \times 
{ \left\{ {\rm Im} \mu, {\rm Im } A_f \right\} \over {\rm max}
( M_{\tilde{f}}, M_{\lambda} ) }~\left( { 1 ~{\rm TeV} \over {\rm max} 
( M_{\tilde{f}}, M_{\lambda} ) } \right)^2~\left( {m_f \over 10~{\rm MeV}}
\right), 
\end{equation}
and one can evade the edm constraints by having small phases for $\mu,
A_{e,u,d}$, or heavy 1st/2nd generation scalar fermions.  However,  this 
would involve enlargement of our model parameter space, since one has to 
consider the sbottom sector as well as the stop sector. Therefore, more 
parameters have to be introduced in principle : $m_{\tilde{b}}^2$ and 
$A_b$ where $A_b$ may be complex like $A_t$.  In order to avoid such 
enlargement, we will assume that there is no accidental cancellation 
between the stop and sbottom loop contributions.  

This CKP edm constraint has not been included in the recent paper by Demir 
{\it et al.} \cite{demir}, who made claims that there could be a large new 
phase shift in the $B^0 - \overline{B^0}$ mixing and it is possible to have 
a fully supersymmetric $\epsilon_K$ from the phases of $\mu$ and $A_t$ only.  
However, if $\tan\beta$ is large ($\tan\beta \approx 60$) as in Ref.~
\cite{demir},  the CKP edm constraints via the CKP mechanism have to be 
properly included. This constraint reduces the possible new phase shift in 
the $B^0 - \overline{B^0}$ mixing to a very small number, $2 | \theta_d | 
\lesssim 1^{\circ}$, as demonstrated in Fig.~1 (a) of Ref.~\cite{baek}.
On the other hand, the CKP edm constraint does not affect too much the 
direct CP asymmetry in $B\rightarrow X_s \gamma$ \cite{baek}.  

In this work, we continue studying the effects of the phases of $\mu$ and 
$A_t$ on $B \rightarrow X_s l^+ l^-$ and $\epsilon_K$. We also reconsider a 
possibility of fully supersymmetric CP violation, namely generating 
$\epsilon_K$ entirely from the phases of $\mu$ and $A_t$ with vanishing KM 
phase ($\delta_{\rm KM} = 0$). Our conclusion is at variance with the claim
made in Ref.~\cite{demir}.

{\bf 2.}
As in Refs.~\cite{baek} \cite{demir}, we assume that the 1st and the 2nd 
family squarks are degenerate and very heavy in order to solve the 
SUSY FCNC/CP problems.  Only the third family squarks can be light enough 
to affect $B$ and $K$ physics.  We also ignore possible flavor changing 
squark mass matrix elements that could generate gluino-mediated flavor 
changing neutral current (FCNC) processes in addition to those effects we 
consider below, relegating the details to the existing literature 
\cite{randall}-\cite{kkl}. 
Therefore the only source of the FCNC in our case is the CKM matrix, 
whereas there are new CPV phases coming from the phases of $\mu$ and $A_t$ 
parameters (see below), in addition to the KM phase $\delta_{KM}$. 
Definitions for the chargino and stop mass matrices are the same as Ref.~
\cite{baek}. There are two new flavor conserving CPV phases in our model, 
${\rm Arg} (\mu)$ and ${\rm Arg} (A_t)$ in the basis where $M_2$ is real. 

We scan over the MSSM parameter space as in Ref.~\cite{baek}
indicated below 
(including that relevant to the EWBGEN scenario in the MSSM) :
\begin{eqnarray}
  \label{eq:input}
  80~{\rm GeV} < | \mu | < 1~{\rm TeV} , &~~& 
  80~{\rm GeV} < M_2 < 1~{\rm TeV} ,
\nonumber   \\
  60~{\rm GeV} < M_A < 1~{\rm TeV} ,
&~~& 2 < \tan\beta < 70,
\nonumber    \\
  (130~{\rm GeV})^2 < M_Q^2 & < & ( 1 ~{\rm TeV} )^2 ,
\nonumber  \\
  - ( 80~{\rm GeV})^2 < M_U^2 & < & (500~{\rm GeV})^2 ,
\nonumber     \\
0 < \phi_{\mu}, \phi_{A_t} < 2 \pi ,&~~&  0 < | A_t | < 1.5 
~{\rm TeV},
\end{eqnarray}
with the following experimental constraints : 
$M_{\tilde{t}_1} > 80$ GeV independent of the mixing angle 
$\theta_{\tilde{t}}$, $M_{\tilde{\chi^{\pm}}} > 83$ GeV, and 
$0.77 \leq R_{\gamma} \leq 1.15$ \cite{alexander}, 
where $R_{\gamma}$ is defined as 
$R_{\gamma}= BR(B \to X_s \gamma)^{expt}/ BR(B \to X_s \gamma)^{SM}$ and 
$BR(B \to X_s \gamma)^{SM} = (3.29 \pm 0.44) \times 10^{-4}$. 
We also impose ${\rm Br} (B\rightarrow X_{sg}) < 6.8 \%$ \cite{bsg}, and 
vary $\tan\beta$ from 2 to 70 \footnote{This may be too large for 
perturbation theory to be valid, but we did extend to $\tan\beta \sim
70$ in order to check the claims made in Ref.~\cite{demir}.}.
This parameter space is larger than that in  the constrained MSSM (CMSSM) 
where the universality of soft terms at the GUT scale is assumed.
Especially, our parameter space includes the electroweak baryogenesis 
scenario in the MSSM  \cite{cline}.  
In the numerical analysis, we used the following numbers for the input 
parameters (running masses in the $\overline{MS}$ scheme are used for 
the quark masses) : $\overline{m_c}(m_c(pole)) = 1.25$ GeV, 
$\overline{m_b}(m_b(pole)) = 4.3$ GeV, $\overline{m_t}(m_t(pole)) = 165$ GeV,
and  $| V_{cb} | = 0.0410, |V_{tb}| = 1, 
| V_{ts} | = 0.0400$ and $\gamma ( \phi_3 ) = 90^{\circ}$ in the CKM matrix 
elements. 

{\bf 3.} 
Let us first consider the branching ratio for $B\rightarrow X_s l^+ l^-$. 
The SM and the MSSM contributions to this decay were considered by several 
groups \cite{bsll} and \cite{bsll_susy}, respectively. We use the standard 
notation for the effective Hamiltonian for this decay as described in Refs.~
\cite{bsll} and \cite{bsll_susy}. The new CPV phases 
in $C_{7,9,10}$ can affect the branching ratio and other observables in 
$B\rightarrow X_s l^+ l^-$ as discussed in the first half of Ref.~\cite{kkl}
in a model independent way. In the second half of Ref.~\cite{kkl}, specific 
supersymmetric models were presented where new CPV phases reside in flavor 
changing squark mass matrices.  In the present work, new CPV phases lie in 
flavor conserving sector, namely in $A_t$ and $\mu$ parameters. 
Although these new phases are flavor conserving, they affect the branching  
ratio of $B\rightarrow X_s l^+ l^-$ and its correlation with $Br (B\rightarrow 
X_s \gamma)$, as discussed in the first half of Ref.~\cite{kkl}.  Note that
$C_{9,10}$ depend on the sneutrino mass, and we have scanned  over 
$ 60 ~{\rm GeV} < m_{\tilde{\nu}} < 200~{\rm GeV}$.
In the numerical evaluation for $R_{ll} \equiv {\rm Br} (B\rightarrow X_s 
l^+ l^-) / {\rm Br} (B\rightarrow X_s l^+ l^-)_{\rm SM}$, we considered 
the nonresonant contributions only for simplicity, neglecting the 
contributions  from $J/\psi, \psi^{'}, etc$.. 
It would be straightforward to incorporate these resonance effects.
In Figs.~1 (a) and (b), we plot the correlations of $R_{\mu\mu}$ with  
${\rm Br} (B\rightarrow X_s \gamma)$ and $\tan\beta$, respectively.
Those points that (do not) satisfy the CKP edm constraints are denoted by 
the squares (crosses). Some points are denoted by both the square and the 
cross. This means that there are two classes of points in the MSSM parameter 
space, and for one class the CKP edm constraints are satisfied but for 
another class the CKP edm constraints are not satisfied, and these two 
classes happen to lead to the same branching ratios for $B\rightarrow X_s 
\gamma$ and $R_{ll}$.  In the presence of the new phases $\phi_{\mu}$ and 
$\phi_{A_t}$, $R_{\mu\mu}$ can be as large as 1.85, and the deviations from 
the SM prediction can be large, if $\tan\beta > 8$.   
As noticed in Ref.~\cite{kkl}, the correlation between the ${\rm Br}
( B\rightarrow X_s \gamma)$ and $R_{ll}$ is distinctly different from that 
in the minimal supergravisty case \cite{okada}. In the latter case, only the 
envelop of Fig.~1 (a) is allowed, whereas everywhere in between is allowed 
in the presence of new CPV phases in the MSSM.  Even if one introduces the 
phases of $\mu$ and $A_0$ at GUT scale in the minimal supergravity scenario, 
this correlation does not change very much from the case of the minimal 
supergravity scenario with real $\mu$ and $A_0$, since the $A_0$ phase becomes 
very small at the electroweak scale because of the renormalization effects 
\cite{keum}. Only $\mu$ phase can affect the  electroweak scale physics, but 
this phase is strongly constrained by the usual edm constraints so that 
$\mu$ should be essentially real parameter. 
Therefore the correlation between $B\rightarrow X_s \gamma$ and $R_{ll}$
can be a clean distinction between the minimal supergravity scenario and 
our model (or some other models with new CPV phases in the flavor changing 
\cite{kkl}).   

{\bf 4.} 
The new complex phases in $\mu$ and $A_t$ will also affect the $K^0 - 
\overline{K^0}$ mixing. The relevant $\Delta S = 2 $ effective Hamiltonian 
is given by 
\begin{equation}
H_{\rm eff}^{\Delta S = 2} = - {G_F^2 M_W^2 \over (2 \pi )^2}~
\sum_{i=1}^3 C_i Q_i,
\end{equation}
where 
\begin{eqnarray}
C_1 ( \mu_0 ) & = & \left( V_{td}^* V_{ts} \right)^2 \left[ 
F_V^W (3;3) + F_V^H (3;3) + A_V^C \right] 
\nonumber  \\
& + & \left( V_{cd}^* V_{cs} \right)^2 \left[ 
F_V^W (2;2) + F_V^H (2;2) \right] 
\nonumber  \\
& + & 2 \left( V_{td}^* V_{ts} V_{cd}^* V_{cs} \right) 
~\left[ F_V^W (3;2) + F_V^H (3;2) \right],
\nonumber  \\
C_2 ( \mu_0 ) & = & \left( V_{td}^* V_{ts} \right)^2 
F_S^H (3;3)  
 + \left( V_{cd}^* V_{cs} \right)^2 ~F_S^H (2;2) 
\nonumber  \\
& + & 2 \left( V_{td}^* V_{ts} V_{cd}^* V_{cs} \right) ~F_S^H (3;2), 
\nonumber  \\
C_3 ( \mu_0 ) & = & \left( V_{td}^* V_{ts} \right)^2 A_S^C,
\end{eqnarray}
where the charm quark contributions have been kept.
The superscripts $W,H,C$ denote the $W^{\pm}, H^{\pm}$ and chargino 
contributions respectively, and  
\begin{eqnarray}
A_V^C & = & \sum_{i,j=1}^2 \sum_{k,l=1}^2
~{1\over 4}~G^{(3,k)i} G^{(3,k)j*} G^{(3,l)i*} G^{(3,l)j} 
Y_1 (r_k, r_l, s_i, s_j ),
\nonumber   \\
A_S^C & = & \sum_{i,j=1}^2 \sum_{k,l=1}^2
~H^{(3,k)i} G^{(3,k)j*} G^{(3,l)i*} H^{(3,l)j} 
Y_2 (r_k, r_l, s_i, s_j ).
\end{eqnarray}
Here $G^{(3,k)i}$ and $H^{(3,k)i}$ are the couplings of $k-$th stop and 
$i-$th chargino with left-handed and right-handed quarks, respectively :
\begin{eqnarray}
G^{(3,k)i} & = & \sqrt{2} C_{R 1i}^* S_{t k1} - 
{ C_{R 2i}^* S_{t k2} \over \sin\beta } ~{m_t \over M_W},
\nonumber \\
H^{(3,k)i} & = & { C_{L 2 i}^* S_{tk1} \over \cos\beta } ~{m_s \over M_W},
\end{eqnarray}
and $C_{L,R}$ and $S_t$ are unitary matrices that diagonalize the chargino
and stop mass matrices \cite{branco}. 
: $C_R^{\dagger} M_{\chi}^- C_L = {\rm diag} 
( M_{\tilde{\chi_1}},M_{\tilde{\chi_2}} )$ and $S_t M_{\tilde{t}}^2
S_t^{\dagger} = {\rm diag} ( M_{\tilde{t}_1}^2, M_{\tilde{t}_2}^2 )$. 
Explicit forms for functions $Y_{1,2}$ and $F$'s can be found in 
Ref.~\cite{branco}, and $r_k = M_{\tilde{t}_k}^2 / M_W^2$ and 
$s_i = M_{\tilde{\chi^{\pm}}_i} / M_W^2$. 
It should be noted that $C_2 ( \mu_0 )$ was misidentified as 
$C_3^H ( \mu_0 )$ in Ref.~\cite{demir}.
The gluino and neutralino contributions are negligible in our model.
The Wilson coefficients at lower scales are obtained by renomalization
group running. The relevant formulae with the NLO QCD corrections at 
$\mu = 2$ GeV are given in  Ref.~\cite{contino}.
It is important to note that $C_1 ( \mu_0 )$ and $C_2 ( \mu_0 )$ are real
relative to the SM contribution in our model. On the other hand, the
chargino exchange contributions to $C_3 (\mu_0 )$ (namely $A_S^C $) are 
generically complex relative to the SM contributions, and  can generate 
a new phase shift in the $K^0 - \overline{K^0}$ mixing relative to the 
SM value. This effect is in fact significant for large 
$\tan\beta (\simeq 1/\cos\beta)$ \cite{demir}, 
since $C_3 (\mu_0)$  is proportional to $ (m_{s} / M_W \cos\beta )^2$. 

The CP violating parameter  $\epsilon_K$ can be calculated from 
\begin{equation}
\epsilon_K \simeq {e^{i \pi / 4}~{\rm Im} M_{12} \over \sqrt{2} \Delta M_K},
\end{equation}
where $M_{12}$ can be obtained from the $\Delta S = 2 $ effective Hamiltonian 
through $2 M_K M_{12} = \langle K^0 | H_{\rm eff}^{\Delta S = 2} | 
\overline{K^0} \rangle$.  For $\Delta M_K$, we use the experimental value 
$\Delta M_K = (3.489 \pm 0.009) \times 10^{-12}$ MeV, instead of theoretical 
relation $\Delta M_K =  2 {\rm Re} M_{12}$,  since the long distance 
contributions to $M_{12}$ is hard to calculate reliably unlike the 
$\Delta S = 2$ box diagrams.  For the strange quark mass, we use the 
$\overline{\rm MS}$ mass at $\mu = 2$ GeV scale : 
$m_s (\mu = 2 {\rm GeV}) = 125$
MeV.  In Figs.~2 (a) and (b), we plot the results of scanning the MSSM 
parameter space : the correlations between $\epsilon_K / \epsilon_K^{\rm SM}$ 
and (a) $\tan\beta$ and (b) the lighter stop mass. We note that $\epsilon_K / 
\epsilon_K^{\rm SM}$ can be as large as $1.4 $  for $\delta_{KM} = 
90^{\circ}$ if $\tan\beta$ is small. This is a factor 2 
larger deviation from the SM compared to the minimal supergravity case 
\cite{kek}. The dependence on the lighter stop is close to the case of the 
minimal supergravity case, but we can have a larger deviations. 
Such deviation is reasonably close to the experimental value, and will 
affect the CKM phenomenology at a certain level. 

In the MSSM with new CPV phases, there is an intriguing possibility that 
the observed CP violation in $K_L \rightarrow \pi\pi$ is fully due to the 
complex parameters $\mu$ and $A_t$ in the soft SUSY breaking terms which 
also break CP softly. 
This possibility was recently considered by Demir {\it et al.}~\cite{demir}.  
Their claim was that it was possible to generate $\epsilon_K$ entirely from 
SUSY CPV phases for large $\tan\beta \approx 60$ with certain choice of
soft parameters 
\footnote{Their choice of parameters leads to $M_{\chi^{\pm}} = 80$ GeV and 
$M_{\tilde{t}} = 85$ GeV, which are very close to the recent lower limits 
set by LEP2  experiments.}. 
In such a scenario, only ${\rm Im}~(A_S^C)$ in Eq. (6) can contribute to 
$\epsilon_K$, if we ignore a possible mixing between $C_2$ and $C_3$ under 
QCD renormalization. In actual numerical analysis we have included this 
effect using the results in Ref.~\cite{contino}.  We repeated their 
calculations using the same set of parameters, but could  not confirm their 
claim. For $\delta_{KM} = 0^{\circ}$, we found that the supersymmetric 
$\epsilon_K$ is less than $\sim 2\times 10^{-5}$, which is too small compared 
to the observed value : $ | \epsilon_K | = (2.280 \pm 0.019) \times 10^{-3}$ 
determined from  $K_{L,S} \rightarrow \pi^+ \pi^-$ \cite{pdg98}.

Let us give a simple estimate for fully supersymmetric $\epsilon_K$, in
which case only $C_3 ( \mu_0 )$ develops imaginary part and can contribute to 
$\epsilon_K$. For $m_{\tilde{t}_1} \sim m_{\chi^{\pm}} \sim M_W$, we would 
get $Y_2 \sim Y_2 (1,1,1,1) = 1/6$, and 
\[
| G^{(3,k) i} | \lesssim O(1),  ~~~{\rm and}~~~ 
| H^{(3,k) i} | \sim {m_s \tan\beta \over M_W}, 
\]
because any components of unitary matrices $C_R$ and $S_t$ are 
$\lesssim O(1)$. Therefore 
${\rm Im} ( A_S^C ) \lesssim O( 10^{-3} )$. 
Now using
\begin{equation}
{\rm Im} (M_{12}) = -{G_F^2 M_W^2 \over (2 \pi )^2} 
f_K^2 M_K 
\left( {M_K \over m_s } \right)^2~{1\over 24}~B_3(\mu)~{\rm Im} (C_3(\mu)),
\end{equation}
and Eq. (9), we get $|\epsilon_K | \lesssim 2 \times 10^{-5}$.

{\bf 7.} 
In conclusion, we extended our previous studies of SUSY CPV phases to 
$B\rightarrow X_s l^+ l^-$ and $\epsilon_K$. Our results can be summarized 
as follows :
\begin{itemize}
\item The branching ratio for $B\rightarrow X_s l^+ l^-$ can be enhanced upto 
$\sim 85 \%$ compared to the SM prediction, and the correlation between 
${\rm Br} (B\rightarrow X_s \gamma)$ and 
${\rm Br} (B\rightarrow X_s l^+ l^-)$ is distinctly different from the 
minimal supergravity scenario (CMSSM) (even with new CP violating phases)
\cite{okada} in the presence of new CP violating phases in $C_{7,8,9}$
as demonstared in model-independent analysis by Kim, Ko and Lee \cite{kkl}.
\item $\epsilon_K / \epsilon_{K}^{SM}$ can be as large as 1.4 for
$\delta_{KM} = 90^{\circ} $. This is the extent to which the new 
phases in $\mu$ and $A_t$ can affect the construction of the unitarity 
triangle through $\epsilon_K$. 
\item Fully supersymmetric CP violation is not possible even for large 
$\tan\beta \sim 60$ and light enough chargino and stop, contrary to the 
claim made in Ref.~\cite{demir}.  With real CKM matrix elements, 
we get very small $|\epsilon_K| \lesssim O(10^{-5})$, 
which is two orders of magnitude smaller than the experimental value.
\end{itemize}
Before closing this paper, we'd like to emphasize that all of our results 
are based on the assumption that there are no new CPV phases in the flavor 
changing sector. Once this assumption is relaxed, then 
gluino-mediated FCNC with additional new CPV phases may play 
important roles,  and many of our results may change \cite{kkl}.  
Within our assumption,  the results presented here and in Ref.~\cite{baek}
are conservative since we did not impose any conditions on the soft SUSY  
breaking terms except that the resulting mass spectra for chargino, stop
and other sparticles satisfy the current lower bounds from LEP and Tevatron.   
More detailed analysis of phenomenological implications of our  works on  
$B_{d(s)}^0 - \overline{B_{d(s)}^0}$ mixing, $B\rightarrow X_{s(d)} \gamma, 
X_{s(d)} l^+ l^-, B_{s(d)}^0 \rightarrow l^+ l^-$ and their direct CP 
asymmetries will be presented elsewhere.

\acknowledgements
The authors wich to thank G.C. Cho for clarifying $O_2$ and $O_3$ in Ref.
~\cite{demir}, 
and A. Ali, A. Grant, A. Pilaftsis and O. Vives for useful communications.  
A part of this work was done while one of the authors (PK) was visiting
Harvard University under the Distinguished Scholar Exchange Program of
Korea Research Foundation.  
This work is supported in part by KOSEF Contract No. 971-0201-002-2, 
KOSEF through Center for Theoretical Physics at Seoul National University, 
Korea Rsearch Foundation Program 1998-015-D00054 (PK), 
and by KOSEF Postdoctoral Fellowship  Program (SB).


%
%


\begin{figure}
\vspace{1cm}
\centerline{\epsfxsize=9.3cm \epsfbox{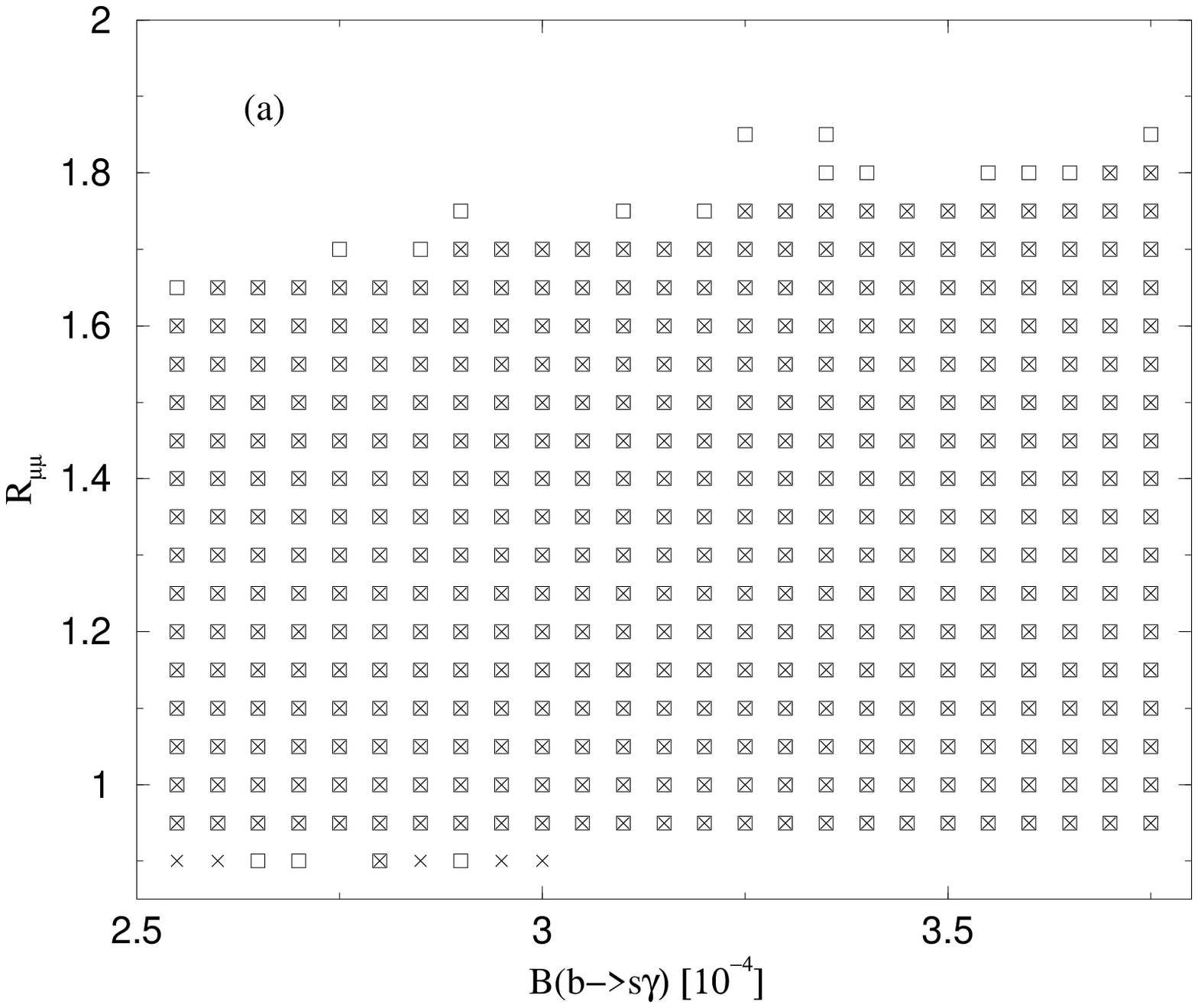}}
\vspace{2cm}
\centerline{\epsfxsize=9.3cm \epsfbox{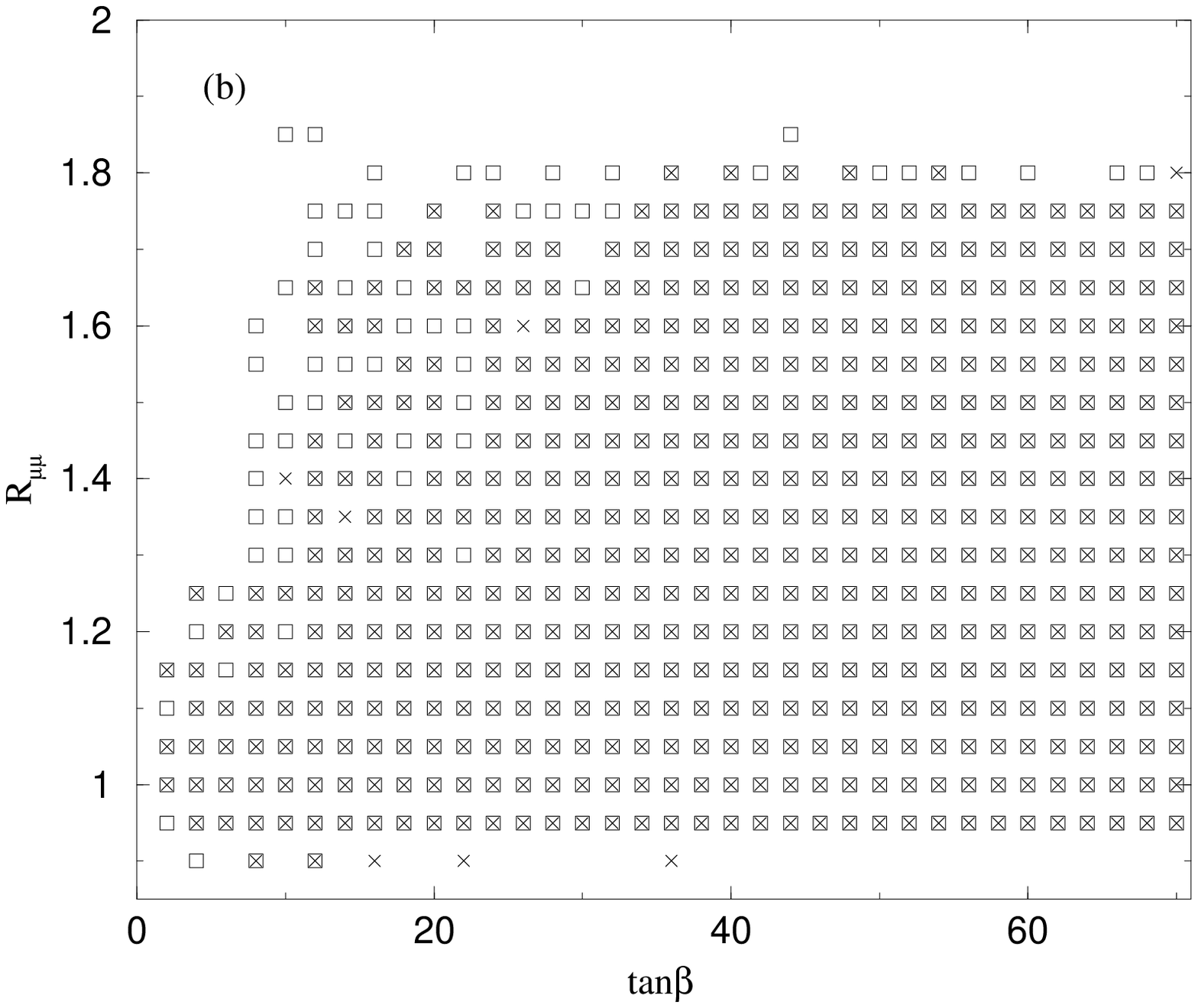}}
\vspace{1cm}
\caption{
The correlations of $R_{\mu\mu}$ with (a) ${\rm Br}(B\rightarrow X_s \gamma)$ 
and (b) $\tan\beta$. 
The squares (the crosses) denote those which (do not) satisfy the CKP edm 
constraints. 
}
\label{fig1}
\end{figure}

\begin{figure}
\vspace{1cm}
\centerline{\epsfxsize=9.3cm \epsfbox{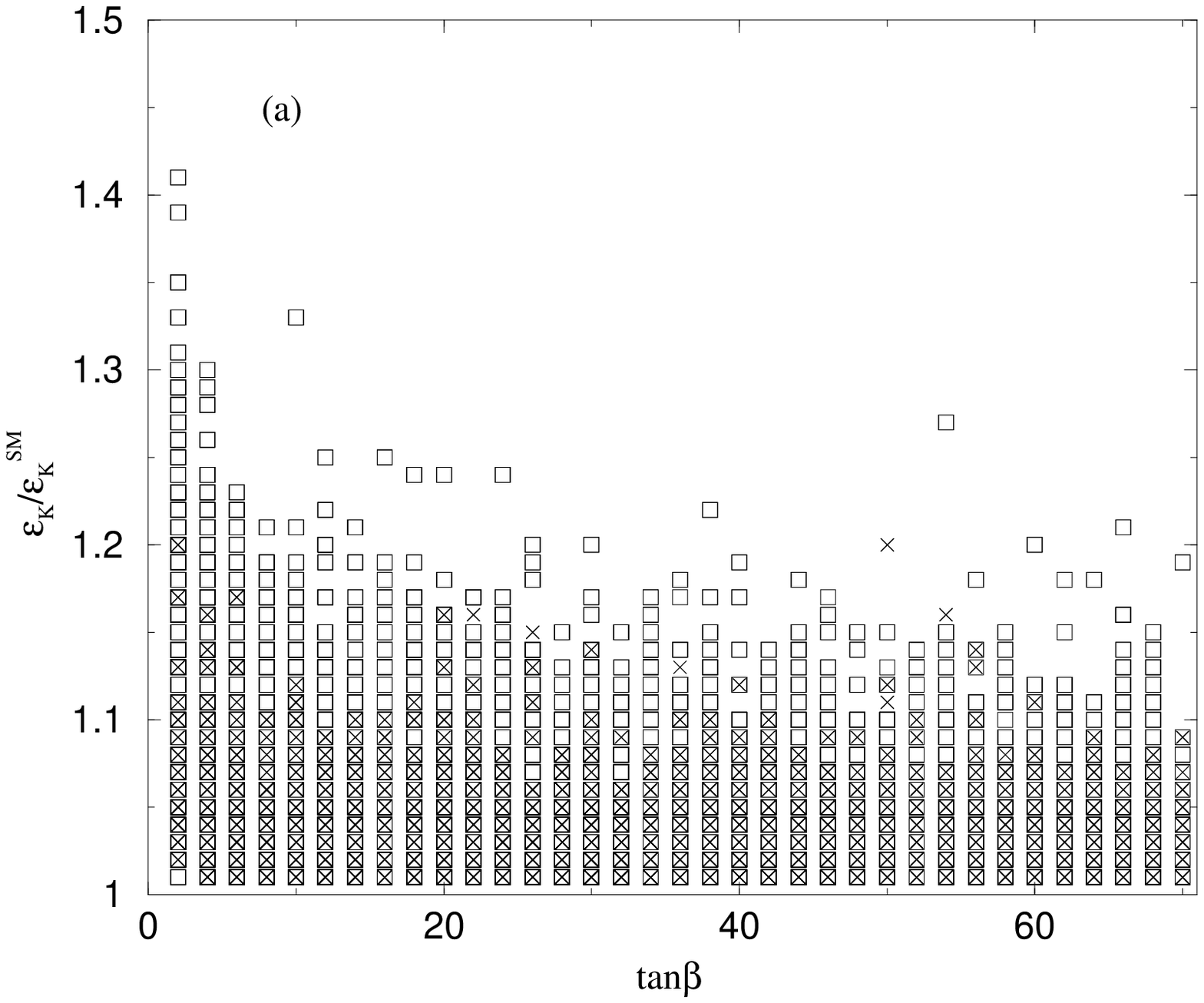}}
\vspace{2cm}
\centerline{\epsfxsize=9.3cm \epsfbox{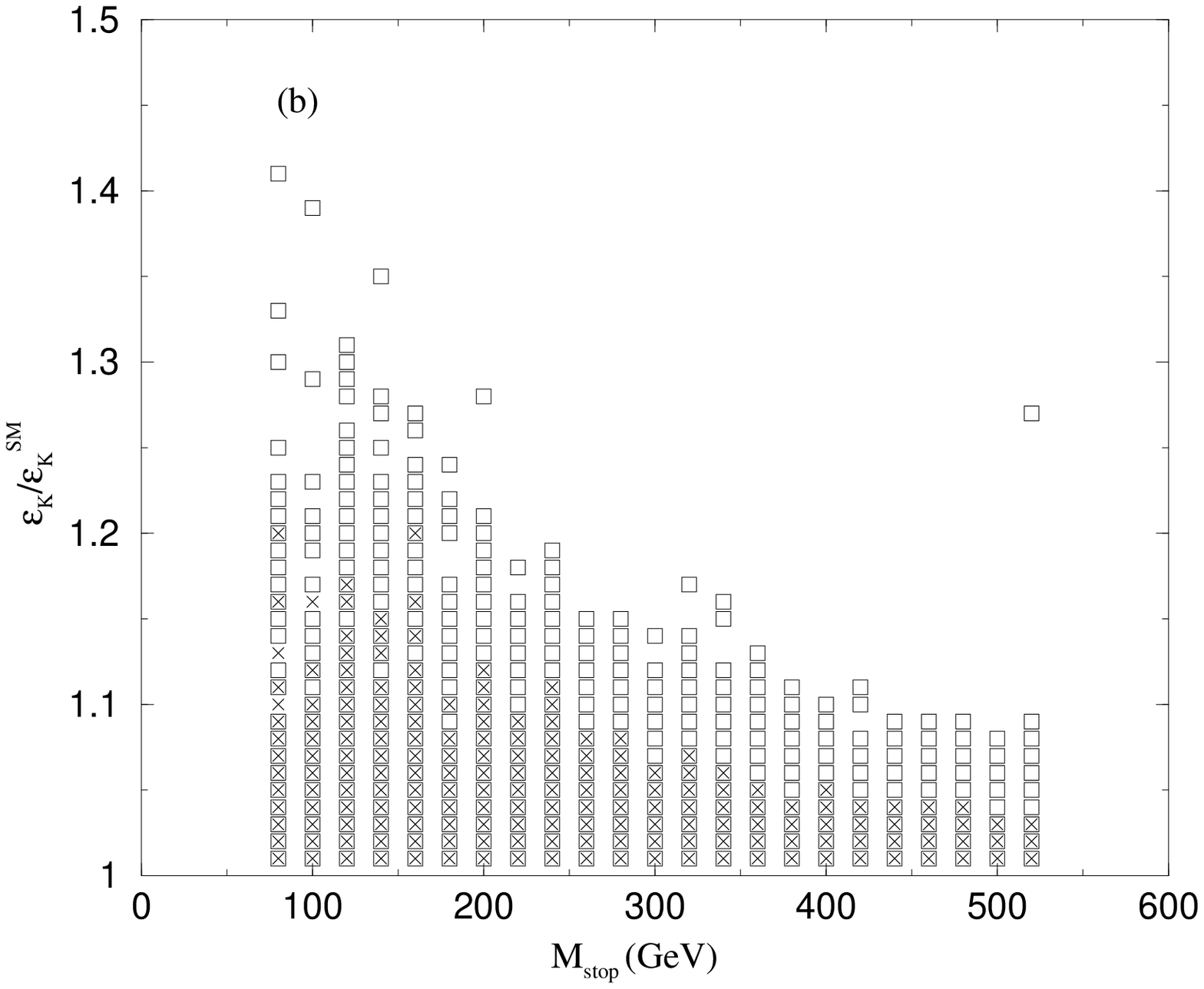}}
\vspace{1cm}
\caption{
The correlations between $\epsilon_K / \epsilon_K^{SM}$ and the lighter 
chargino mass $M_{\tilde{\chi_1^{\pm}}}$ for (a) $2< \tan\beta < 35$  and 
(b) $35 < \tan\beta < 70$, respectively. 
The squares (the crosses) denote those which (do not) satisfy the CKP edm 
constraints. 
}
\label{fig2}
\end{figure}


%


\begin{references}
\bibitem{susycp} See, for example, S.M. Barr and W.J. Marciano,
in {\it CP Violation}, edited by C. Jarlskog (World Scientific, Singapore,
1989), p. 455 ; W. Bernreuther and M. Suzuki, Rev. Mod. Phys. {\bf 63},
313 (1991).
\bibitem{kaplan} A.G. Cohen, D.B. Kaplan, A.E. Nelson, Phys. Lett. 
{\bf B388}, 588 (1996) ; A. G. Cohen, David B. Kaplan, F. Lepeintre, 
Ann E. Nelson, Phys. Rev. Lett.{\bf 78}, 2300 (1997).
\bibitem{nath} T. Ibrahim and P. Nath, Phys. Lett. {\bf B 418}, 98 (1998) ;
Phys. Rev. {\bf D 57}, 478 (1998) ; (E) {\it ibid.}, {\bf D 58}, 019901 
(1998) ; Phys. Rev. {\bf D 58}, 111301 (1998).
\bibitem{kane} M. Brhlik, G.J. Good and G.L. Kane, hep-ph/9810457. 
\bibitem{baek} Seungwon Baek and P. Ko, KAIST-20/98, SNUTP 98-139,
hep-ph/9812229 (1998).
\bibitem{demir} D.A. Demir, A. Masiero and O. Vives, 
Phys. Rev. Lett. {\bf 82}, 2447 (1999).
\bibitem{chang} D. Chang, Wai-Yee Keung, Apostolos Pilaftsis,
Phys. Rev. Lett. {\bf 82}, 900 (1999).  
\bibitem{randall} L. Randall and S. Su, Nucl.Phys. {\bf B 540}, 37 (1999).
\bibitem{hou} C.-K. Chua, X.-G. He and W.-S. Hou, hep-ph/9808431.
\bibitem{kkl} Y.G. Kim, P. Ko and J.S. Lee, Nucl. Phys. {\bf B 544}, 64 (1999).
\bibitem{alexander} J. Alexander, plenary talk at ICHEP98, Vancouver, Canada.
\bibitem{bsg} T.E. Coan {\it et al.} (CLEO Collaboration), Preprint 
CLNS 97/1516, Phys. Rev. Lett. {\bf 80}, 1150 (1998).
\bibitem{cline} M. Carena and C.E.M. Wagner, hep-ph/9704347 ; 
J. M. Cline, M. Joyce and K. Kainulainen, Phys. Lett.{\bf B 417}, 79 (1998) ;
M. Carena, M. Quiros and C.E.M. Wagner, Nucl.Phys. {\bf B 524}, 3 (1998) ;
J.M. Cline and G.D. Moore, Phys. Rev.Lett. {\bf 81}, 3315 (1998).
\bibitem{bsll} B. Grinstein, M.J. Savage and M.B. Wise, Nucl. Phys. 
{\bf B 319}, 271 (1994) ; M. Misiak, Nucl. Phys. {\bf B 393} 23 (E) (1993) ;
{\bf 439}, 461 (1995) ; 
A.J. Buras and M. M\"{u}nz, Phys. Rev. {\bf D 52}, 186 (1995).
\bibitem{bsll_susy} S. Bertolini, F. Borzumati, A. Masiero and G. Ridolfi,
Nucl. Phys. {\bf 353}, 591 (1991) ; 
P. Cho, M. Misiak and D. Wyler, Phys. Rev. {\bf D 54}, 3329 (1996).
\bibitem{okada} T. Goto, Y. Okada and Y. Shimizu,
Phys. Rev. {\bf D 58}, 094006 (1998).
\bibitem{keum} T. Falk and K. Olive,  Phys. Lett. {\bf B 439}, 71 (1998) ; 
T. Goto, Y.Y. Keum, T. Nihei, Y. Okada, Y. Shimizu, hep-ph/9812369.
\bibitem{branco} G.C. Branco, G.C. Cho, Y. Kizukuri and N. Oshimo, 
Phys. Lett. {\bf B 337}, 316 (1994) ; Nucl. Phys. {\bf B 449}, 483 (1995).
\bibitem{contino} R. Contino and I. Scimemi, hep-ph/9809437.
\bibitem{kek} T. Goto, T. Nihei and Y. Okada, Phys. Rev. {\bf D 53}, 5233 
(1996).
\bibitem{pdg98} Particle Data Group, Eur. Phys. J. {\bf C 3}, 1 (1998).
\end{references}
\end{document}